\documentclass[sigconf]{acmart}

\AtBeginDocument{%
  \providecommand\BibTeX{{%
    \normalfont B\kern-0.5em{\scshape i\kern-0.25em b}\kern-0.8em\TeX}}}

\copyrightyear{2022} 
\acmYear{2022} 
\setcopyright{acmcopyright}\acmConference[SIGIR '22]{Proceedings of the 45th International ACM SIGIR Conference on Research and Development in Information Retrieval}{July 11--15, 2022}{Madrid, Spain}
\acmBooktitle{Proceedings of the 45th International ACM SIGIR Conference on Research and Development in Information Retrieval (SIGIR '22), July 11--15, 2022, Madrid, Spain}
\acmPrice{15.00}
\acmDOI{10.1145/3477495.3531718}
\acmISBN{978-1-4503-8732-3/22/07}

\usepackage{multicol}
\usepackage{multirow}
\usepackage{subfig}
\usepackage{arydshln}
\usepackage{acronym}
\usepackage{xspace}
\usepackage{enumitem}
\usepackage{adjustbox}
\newtheorem{definition}{Definition}

\definecolor{blue(pigment)}{rgb}{0.2, 0.2, 0.6}

\newcommand{\model}{UFR\xspace}

\newcommand{\eg}{e.g., }
\newcommand{\ie}{i.e., }

\acrodef{RS}{recommender systems}

\newcommand{\partitle}[1]{\vspace{2mm}\noindent\textbf{#1}}

\begin{document}

\title[Experiments on Generalizability of User-Oriented Fairness in Recommender Systems]{Experiments on Generalizability of User-Oriented Fairness \\ in Recommender Systems}

\author{Hossein A.~Rahmani}
\affiliation{%
  \institution{University College London}
  \city{London}
  \country{United Kingdom}
}
\email{h.rahmani@ucl.ac.uk}

\author{Mohammadmehdi Naghiaei}
\affiliation{%
  \institution{University of Southern California}
  \city{California}
  \country{USA}
}
\email{naghiaei@usc.edu}

\author{Mahdi Dehghan}
\affiliation{%
  \institution{Shahid Beheshti University}
  \city{Tehran}
  \country{Iran}
}
\email{mahdi.dehghan551@gmail.com}

\author{Mohammad Aliannejadi}
\affiliation{%
  \institution{University of Amsterdam}
  \city{Amsterdam}
  \country{The Netherlands}
}
\email{m.aliannejadi@uva.nl}

\renewcommand{\shortauthors}{H.~A.~Rahmani, M.~Naghiaei, M.~Dehghan, M.~Aliannejadi}

\begin{abstract}
Recent work in recommender systems mainly focuses on fairness
in recommendations as an important aspect of measuring recommendations quality. A fairness-aware recommender system aims to treat different user groups similarly. Relevant work on user-oriented fairness highlights the discriminative behavior of fairness-unaware recommendation algorithms towards a certain user group, defined based on users' activity level. Typical solutions include proposing a user-centered fairness re-ranking framework applied on top of a base ranking model to mitigate its unfair behavior towards a certain user group \ie disadvantaged group. In this paper, we re-produce a user-oriented fairness study and provide extensive experiments to analyze the dependency of their proposed method on various fairness and recommendation aspects, including the recommendation domain, nature of the base ranking model, and user grouping method. Moreover, we evaluate the final recommendations provided by the re-ranking framework from both user- (\eg NDCG, user-fairness) and item-side (\eg novelty, item-fairness) metrics. We discover interesting trends and trade-offs between the model's performance in terms of different evaluation metrics. For instance, we see that the definition of the advantaged/disadvantaged user groups plays a crucial role in the effectiveness of the fairness algorithm and how it improves the performance of specific base ranking models. Finally, we highlight some important open challenges and future directions in this field. We release the data, evaluation pipeline, and the trained models publicly on \url{https://github.com/rahmanidashti/FairRecSys}.
\end{abstract}



\maketitle

\section{Introduction and Context}
\label{sec:intro}

\Ac{RS} are mainly aimed to enhance recommendation effectiveness for users based on their past interactions, hence being evaluated by typical ranking-based metrics. Much research argues that such an approach fails to take into account critical aspects of recommendation, such as fairness in one- and two-sided marketplaces~\cite{melchiorre2020personality, beutel2019fairness, burke2018balanced}, as well as biased behavior of algorithms towards certain groups of items and users~\cite{ekstrand2018all, melchiorre2020personality, li2021user, abdollahpouri2019unfairness}. Therefore, one line of research seeks to propose fair and unbiased algorithms from various aspects~\cite{abdollahpouri2017controlling, Chen2021autodebias, Lin2021mitigating}, while another line of research looks at various biases in the data and how they may affect algorithms trained on them, leading to an unfair behavior of the system~\cite{abdollahpouri2019unfairness,neophytou2021revisiting,abdollahpouri2021user}. Examples of such biases are position~\cite{wu2021unbiased,zhao2019recommending} and trust~\cite{agarwal2019addressing} bias in recommendation.

Various works have studied the fairness of recommender systems under different assumptions such as group and individual fairness. Group fairness concerns the capability of generating equitable recommendations to different types of users, particularly differentiated depending on sensitive attributes \eg gender, age, personality, and race~\cite{ekstrand2018all, melchiorre2020personality}.
Such studies involve two primary perspectives, namely, whether the market is one-sided or two-sided. Fairness on one-sided markets~\cite{li2021towards,hao2021pareto} aims at minimizing the disparity between different \textit{user} groups and removing the algorithm's bias against the ``protected'' user group in the recommendation process. 
In contrast, fairness on two-sided markets~\cite{DBLP:conf/ictir/WangJ21,DBLP:conf/kdd/SuhrBZGC19,wu2021tfrom,sarvi2022wsdm} aims to protect not only the protected users, but also some item groups, leading to a fairer approach towards certain content providers (\eg music and video recommendation) or vendors (\eg eCommerce recommendation). 

Methods on fair ranking can be categorized into three groups, namely, pre-process, in-process, and post-process. Pre-processing approaches are formed on the basis that often unfairness issues is caused by the biased, discriminatory, or imbalanced distributions of sensitive features~\cite{abdollahpouri2017controlling}, and are focused on removing bias and unfairness in data before training the model. The second category directly learns a fair ranking model~\cite{narasimhan2020pairwise, singh2019policy, zehlike2020reducing}, whereas the third one considers re-ranking and post-processing the output of a utility-optimized ranking~\cite{biega2018equity, celis2017ranking, li2021user}. Post-processing fairness ranking approaches have attracted lots of interest in the community because they provide a tool that can transform various state-of-the-art black-box ranking methods to a fair ranking. Moreover, such techniques are practical choices in many use cases since they do not require re-training in case the definition of fairness or protected groups changes. This is a critical feature, especially when group and fairness definitions are dynamic and re-training a model comes at a high cost.

As mentioned earlier, various bias and fairness aspects (\eg item and user) can influence how one interprets a fair ranking. Moreover, pre- and post-processing fairness approaches introduce a new dimension, \ie the base ranking model. We observe that various research studies provide post-processing solutions, applied to a limited number of ranking models, on a limited number of domains (\ie datasets). Moreover, in the majority of the current studies, only a limited number of fairness aspects are considered. For instance, \citet{ferraro2021break} propose a re-ranking algorithm for user-oriented fairness, where they only evaluate the performance of their method on one type of bias and domain \ie gender and music, respectively. Moreover, they conduct their experiments on a limited range of base ranking models including collaborative filtering approaches and the most famous baselines such as MostPop. \citet{liu2019personalized} also propose a re-ranking fairness algorithm on a dataset generated from Kiva.org. Although they consider fairness in two-sided markets, they only apply their proposed method to one domain. Moreover, they do not test the performance of their method when applied on top of a wide range of base ranking models (\eg deep models are ignored in their experiments).
Experimenting with post-processing methods against multiple base ranking models is crucial, especially when the fairness algorithm does not assume having access to the relevance judgments, since the distribution of top-ranked items and the model's learned estimate of relevance can influence the performances to a great extent. 

To this aim, in this paper, we re-implement and reproduce the results of a state-of-the-art user fairness post-processing framework on \textit{eight} recommendation datasets, consisting of \textit{six} domains. In particular, we provide an in-depth experimental analysis of \citet{li2021user}, based on several reproducibility aspects, namely, (i) domain, (ii) base recommendation algorithm, (iii) user group assumptions, and (iv) interplays in other effectiveness metrics including item fairness. In summary, our reproducibility study concerns the following aspects:

\begin{itemize}[leftmargin=*]
    \item We conduct extensive experiments on various domains with different feedback types (\ie explicit and implicit) and characteristics, namely, MovieLens, Epinions, BookCrossing, AmazonToy, AmazonOffice, Gowalla, Foursquare, and Last.fm. We investigate whether the user-oriented fairness re-ranking (\model) method can be applied to diverse domains and whether underlying data characteristics (\eg sparsity and average user interactions) can explain the degree of improvement in fairness.
    \item We examine the effectiveness of \model on \textit{six} new models, including MostPop \cite{dacrema2019we}, BPR \cite{rendle2012bpr}, PF \cite{gopalan2015scalable}, WMF \cite{hu2008collaborative}, NeuMF \cite{he2017neural}, and VAECF \cite{liang2018variational}. We aim to analyze how different baseline algorithms influence the fairness performance in the proposed model. In particular, we investigate whether some baseline algorithms (\eg shallow vs.~deep learning-based models) are less prone to unfair treatment between user groups.
    \item We further analyze the impact of improving user-oriented fairness in two-sided markets. In particular, we investigate whether improving fairness from the user's perspective affects the producer's visibility, system catalog coverage, and existing popularity bias, \ie over-emphasizing popular items (\ie short-head items) with limited visibility to less popular items (\ie long-tail items). 
    \item In addition, we investigate the underlying relation between user-oriented fairness and user-oriented evaluation of popularity bias, known as popularity bias calibration fairness~\cite{abdollahpouri2019unfairness,naghiaei2022unfairness}, \ie the degree to which popularity bias impacts users with different levels of interest in popular items in their profiles.
    \item Finally, we investigate the effectiveness of \model under different user groups based on their activity level and interaction with popular items. We analyze the user grouping effect on the improvement of fairness. We further study whether dividing users into groups based on a cutoff (\eg top $5\%$) can create two categories of (dis)advantaged users in a meaningful way across various domains.
\end{itemize}

Our reproducibility experiments on \model lead to critical observations and implications on this problem. For instance, our experiments on different domains indicate that the effectiveness of \model highly depends on the domain. Also, we found a high dependency of fairness performance on the way advantaged/disadvantaged users are defined. This study provides important guidelines on how to evaluate fairness methods more effectively, particularly when proposing a fairness algorithm that involves only one of multiple stakeholders (\ie only the user or the vendor) since it can affect the exposure of other stakeholders. Therefore, it is crucial for a wide range of fairness algorithms to be evaluated thoroughly, following our proposed guidelines.
\section{Reproducibility Methodology}
\label{sec:method}
In this section, we first describe the approach that we reproduce. Then, we provide the details of the datasets and the recommendation algorithms used in the fairness re-ranking algorithm. Finally, we give a detailed description on how we define the advantaged and disadvantaged users groups.

\subsection{\model Implementation}
\citet{li2021user} consider user unfairness across different user groups based on the level of activity in the platform (the more or the less active). They adopt a constrained form of re-ranking integer programming, as opposed to \citet{Naghiaei2022CPFairPC} that used an unconstrained version. The user-oriented fairness re-ranking algorithm (UFR) objective function is to select (or re-rank) the recommendation list of each user provided by the baseline algorithms to maximize the total score (accuracy) while constrained by keeping the difference in average recommendation performance between the groups of users in a given range. They use F1@10 and NDCG@10 to assess recommendation performance quality and the difference of NDCG between the two user groups as the constraint, also called user-oriented group fairness (UGF). Their evaluation on Amazon datasets shows that their re-ranking approach leads to improved fairness between the two user groups significantly, while at the same time improving the overall recommendation utility. The original code is available in a public GitHub repository\footnote{\url{https://github.com/rutgerswiselab/user-fairness}}, however, it contains only the fairness-aware post-processing re-ranking method and lacks the training and testing pipeline of the baseline models used in the study. Moreover, they make use of Gurobi\footnote{\url{https://www.gurobi.com/}}, a commercial optimization solver to be able to tune the codes. 

Here, we implement the post-processing pipeline and the integer programming optimization code. In contrast to \citet{li2021user}, we use Cornac\footnote{\url{https://cornac.preferred.ai/}}, a Python-based open-source recommendation toolkit, to provide various state-of-the-art recommendation models, straightforward evaluation pipeline, and to foster the reproducibility of the study. Furthermore, we implement a fairness-aware re-ranking method using MIP\footnote{\url{https://www.python-mip.com/}}, a Python tool for the modeling and optimization, on top of Groubi, enabling us to run the code with a license obtained for academic use only. With this setup, we examine $96$ experimental cases on eight well-known recommendation datasets, six state-of-the-art recommendation algorithms, and two user grouping unfairness criteria \ie $8 \ \text{(datasets)} \times \ 6 \ \text{(CF baselines)} \times \ 2 \ (\text{fairness constraints type})$.

\subsection{Datasets}
We use eight public datasets from different domains such as Movie, Point-of-Interest (POI), Music, eCommerce, Book, and Opinion with different types of feedback (\ie explicit and implicit) and characteristics, including MovieLens \cite{harper2015movielens}, Epinions \cite{massa2007trust}, BookCrossing \cite{ziegler2005improving}, AmazonToy \cite{he2016ups}, AmazonOffice \cite{mcauley2015image}, Gowalla \cite{liu2017experimental}, Last.fm\footnote{\url{http://www.lastfm.com}} \cite{Cantador:RecSys2011}, and Foursquare \cite{liu2017experimental}. We apply $k$-core pre-processing (\ie each training user/item has at least $k$ ratings) on the datasets to make sure each user/item has sufficient feedback. Table \ref{tbl:datasets} shows the statistics of all the datasets.

\begin{table*}
  \caption{Statistics of the datasets: $\left| \mathcal{U} \right|$ is the number of users, $\left| \mathcal{I} \right|$ is the number of items, $\left| \mathcal{P} \right|$ is the number of interactions, $\left| \mathcal{SI} \right|$ is the number of short-head items, $\left| \mathcal{LI} \right|$ is the number of long-tail items, $\left| \mathcal{AU} \right|$ and $\left| \mathcal{DU} \right|$ are the number of advantaged and disadvantaged users, respectively, in G1 and G2 (in brackets are the \# of relevant items).}
  \centering
  \label{tbl:datasets}
  \begin{adjustbox}{max width=\textwidth}
  \begin{tabular}{lllllllllllccccc}
    \toprule
    & \multirow{2}{*}{\textbf{Dataset}} & \multirow{2}{*}{$\left| \mathcal{U} \right|$} & \multirow{2}{*}{$\left| \mathcal{I} \right|$} & \multirow{2}{*}{$\left| \mathcal{P} \right|$} & \multirow{2}{*}{$\frac{\left| \mathcal{P} \right|}{\left| \mathcal{U} \right|}$} & \multirow{2}{*}{$\frac{\left| \mathcal{P} \right|}{\left| \mathcal{I} \right|}$} & \multirow{2}{*}{\%Sparsity} & \multirow{2}{*}{Domain} & \multirow{2}{*}{$\left| \mathcal{SI} \right|$} & \multirow{2}{*}{$\left| \mathcal{LI} \right|$} & \multicolumn{2}{c}{G1} && \multicolumn{2}{c}{G2}\\
    \cmidrule{12-13} \cmidrule{15-16}
    & & & & & & & & & & & $\left| \mathcal{AU} \right|$ & $\left| \mathcal{DU} \right|$ && $\left| \mathcal{AU} \right|$ & $\left| \mathcal{DU} \right|$\\
     
    \midrule
    \multirow{5}{*}{\rotatebox[origin=c]{90}{\textbf{Explicit}}} & \textbf{Epinions} & 2,677 & 2,060 & 103,567 & 38.6 & 50.2 & 98.12\% & Opinion & 412 & 1,648 & 134 (12220) & 2542 (60276) && 535 (28653) & 2141 (43843) \\
    &\textbf{MovieLens} & 943 & 1,349 & 99,287 & 105.2 & 73.6 & 92.19\% & Movie & 269 & 1,080 & 47 (12698) & 895 (56802) && 188 (34829) & 754 (34671) \\
    &\textbf{BookCrossing} & 1,136 & 1,019 & 20,522 & 18.0 & 20.1 & 98.22\% & Book & 203 & 816 & 57 (1831) & 1078 (12533) && 227 (4904) & 908 (9460) \\
    &\textbf{AmazonToy} & 2,170 & 1,733 & 32,852 & 15.1 & 18.9 & 99.12\% & eComm. & 346 & 1,387 & 108 (3867) & 2061 (19128) && 434 (9577) & 1735 (13418) \\
    &\textbf{AmazonOffice} & 2,448 & 1,596 & 36,841 & 15.0 & 23.0 & 99.05\% & eComm. & 319 & 1,276 & 122 (3945) & 2325 (21842) && 489 (10392) & 1958 (15395) \\
    \midrule
    \multirow{3}{*}{\rotatebox[origin=c]{90}{\textbf{Implicit}}} &\textbf{Gowalla} & 1,130 & 1,189 & 66,245 & 58.6 & 55.7 & 95.06\% & POI & 237 & 952 & 56 (6452) & 1073 (39919) && 226 (19267) & 903 (27104) \\
    &\textbf{Foursquare} & 1,568 & 1,461 & 42,678 & 27.2 & 29.2 & 98.13\% & POI & 292 & 1169 & 78 (2899) & 1489 (26975) && 313 (10027) & 1254 (19847) \\
    &\textbf{Last.fm} & 1,797 & 1,507 & 62,376 & 34.7 & 41.3 & 97.69\% & Music & 301 & 1,206 & 90 (2447) & 1706 (41215) && 359 (11789) & 1437 (31873) \\
    \bottomrule 
  \end{tabular}
  \end{adjustbox}
\end{table*}

\subsection{Base Ranking Models} 
We compare the performance of \model on several recommendation approaches, from traditional to deep recommendation models, as suggested by \citet{dacrema2019we}. Therefore, we include two traditional methods (PF and WMF), as well as two deep recommendation models (NeuMF and VAECF). We also include two baselines approaches, MostPop and BPR, for further investigation and comparison of the results. Below, we explain the methods in more detail:
\begin{itemize}[leftmargin=*]
    \item \textbf{MostPop} \cite{dacrema2019we}: This method is a non-personalized method that recommends the relevant items to each user. Popularity is measured by the number of interactions of items.
    \item \textbf{BPR} \cite{rendle2012bpr}: This is a state-of-the-art method for personalized ranking, particularly on implicit feedback datasets. BPR uses a pairwise ranking loss whose goal is to optimize personalized ranking.
    \item \textbf{PF} \cite{gopalan2015scalable}: This method is a variant of probabilistic matrix factorization where the weights of each user and item latent features are positive and modeled using the Poisson distribution.
    \item \textbf{WMF} \cite{hu2008collaborative}: This method is a weighted matrix factorization with $l2$-norm regularization that considers the latent features of two items independently.
    \item \textbf{NeuMF} \cite{he2017neural}: This algorithm learns user and item features using the multi-layer perceptron (MLP) on one side and matrix factorization (MF) from another side, then applies non-linear activation functions to train the mapping between users and items features that are concatenated from MLP and MF layers.
    \item \textbf{VAECF} \cite{liang2018variational}: This method is based on variational autoencoders which introduce a generative model with multinomial likelihood and use Bayesian inference for parameter estimation.
\end{itemize}
Note that we use the implementation of all these methods publicly available in the Cornac library. We compare the fair re-ranking algorithm applied on top of the baselines to show how the re-ranking method can achieve the desirable performance on fairness metrics and overall recommendation performance (accuracy and beyond-accuracy).

\partitle{Hyperparameter Setting.}
We adopt the baseline algorithms with the default parameter settings, suggested in their original paper. We set the embedding size for users and items to $50$ for all baseline algorithms. For NeuMF, we set the size of the MLP with $32, 16, 8$ and we apply the hyperbolic tangent (TanH) non-linear activation function between layers. We set the learning rate to $0.001$. We apply Adam \cite{kingma2014adam} as the optimization algorithm to update the model parameters.

\subsection{Fairness Assumption and Evaluation}
\label{sec:fairness_assumption}

In group fairness, the protected groups are treated similarly to the advantaged group \cite{pedreschi2009measuring}. The group of users can be divided under different requirements for different tasks \cite{abdollahpouri2019unfairness,li2021user}. In this study, we examine two different user grouping methods according to the level of activity of users, as well as consumption of popular items in the user profiles. We label the more active users as an advantaged group, while the remaining users as a disadvantaged group as follows:

\begin{definition}
    \textbf{The level of activity (G1):} The study of \citet{li2021user} groups users into advantaged and disadvantaged groups according to their level of activity \ie they select the top 5\% of users in the training dataset ranked by the number of interactions and label them as the advantaged group and the remaining users are labeled as the disadvantaged group. We apply this grouping strategy to our study to highlight and evaluate the same strategy as the study of \citet{li2021user}.
\end{definition}

\begin{definition}
    \textbf{The consumption of popular items (G2)}: Following \cite{abdollahpouri2019unfairness}, we also divide users into advantaged and disadvantaged groups based on the number of popular items in their profiles. We define an item to be popular if its popularity value (\ie the number of interactions) falls within the top 20\% of item popularity. We then select the top 20\% of users in the training dataset ranked by the number of popular items and label them as the advantaged group and the remaining users are labeled as the disadvantaged group.
\end{definition}

The reason for exploring these two methods is that we believe the difference between user interactions and popular consumption will reflect their different activity level in a reasonable manner. In the paper, we refer to the first grouping method as the \textbf{G1} and the second grouping method as the \textbf{G2}.

We evaluate the popularity bias of different user groups (\ie advantaged vs.~disadvantaged) using the \textit{delta Group Average Popularity} ($\Delta$GAP) metric proposed by \citet{abdollahpouri2019unfairness}. For each recommendation algorithm and user group, $\Delta$GAP measures the difference between the popularity of the recommended items from the expected popularity of the items in the user profiles as $\Delta GAP=\frac{GAP(g)_r-GAP(g)_p}{GAP(g)_p}$,
where $g$ is a given user group, $GAP(g)_r$ and $GAP(g)_c$ represent the GAP value for user profiles and recommendations, respectively, and it is defined as $    GAP(g)=\frac{\sum_{u\in{g}}\frac{\sum_{i\in{p_u}}{\phi{(i)}}}{|p_u|}^{}}{|g|}$, 
where $\phi{(i)}$ is the popularity of item $i$, which is the number of times it is rated and $p_u$ is the list of items in the profile of user $u$. The value of $\Delta$GAP = 0 indicates the recommendation is fair, that is the average popularity of recommended items matches the average popularity of the user profile.

\subsection{Relevance Estimation Assumption}
Various post-processing fairness algorithms assume having access to the relevance judgments~\cite{li2021user,wang2021user} in the post-processing step. Therefore, the fairness algorithm is aware of the relevance of the items in the post-processing step and can leverage this information in the re-ranking process. However, in this work, we conduct our experiments with the assumption that the relevance labels are not available to the post-processing algorithm and are estimated based on the training data. Not only is this a more realistic assumption, but it also leads to a more concrete reproducibility study. In particular, studying the effect of the base ranking model is justified mainly in cases where the fairness algorithm takes as input the base ranking and estimates the relevance of the items based on the model's predictions (rather than their true labels). 

\subsection{Code and Data}
We release the code of the complete pipeline, including training and inference of the base ranking models, as well as the fairness-aware re-ranking stage, together with the pre-processed data used to produce the results in this paper. Moreover, we provide ready-to-run Jupyter Notebooks, tested on Google Colab that can be executed to obtain the results and plots reported in the paper. All the related resources can be accessed on our website: \url{https://github.com/rahmanidashti/FairRecSys}

\section{Reproducibility Aspects}
\label{sec:rep.aspects}
In this section, we explain the experiments we conduct on various reproducibility aspects and discuss the obtained results and our observations. Note that, in the interest of space, we only report the detailed results on two datasets, but provide the results on all the datasets on our repository.

\subsection{Domain}
In this experiment, we aim to study the generalizability of \model on different domains. We study the robustness of \model, applied on \textit{eight} different datasets, consisting of \textit{six} domains. Moreover, we explore the effectiveness of different data characteristics including average interactions per user/item, feedback type, sparsity, etc on the capability of \model to mitigate bias towards a certain user group.

\begin{figure}%
    \centering
    \subfloat[User grouping: G1]
    {
        \includegraphics[scale=0.32]{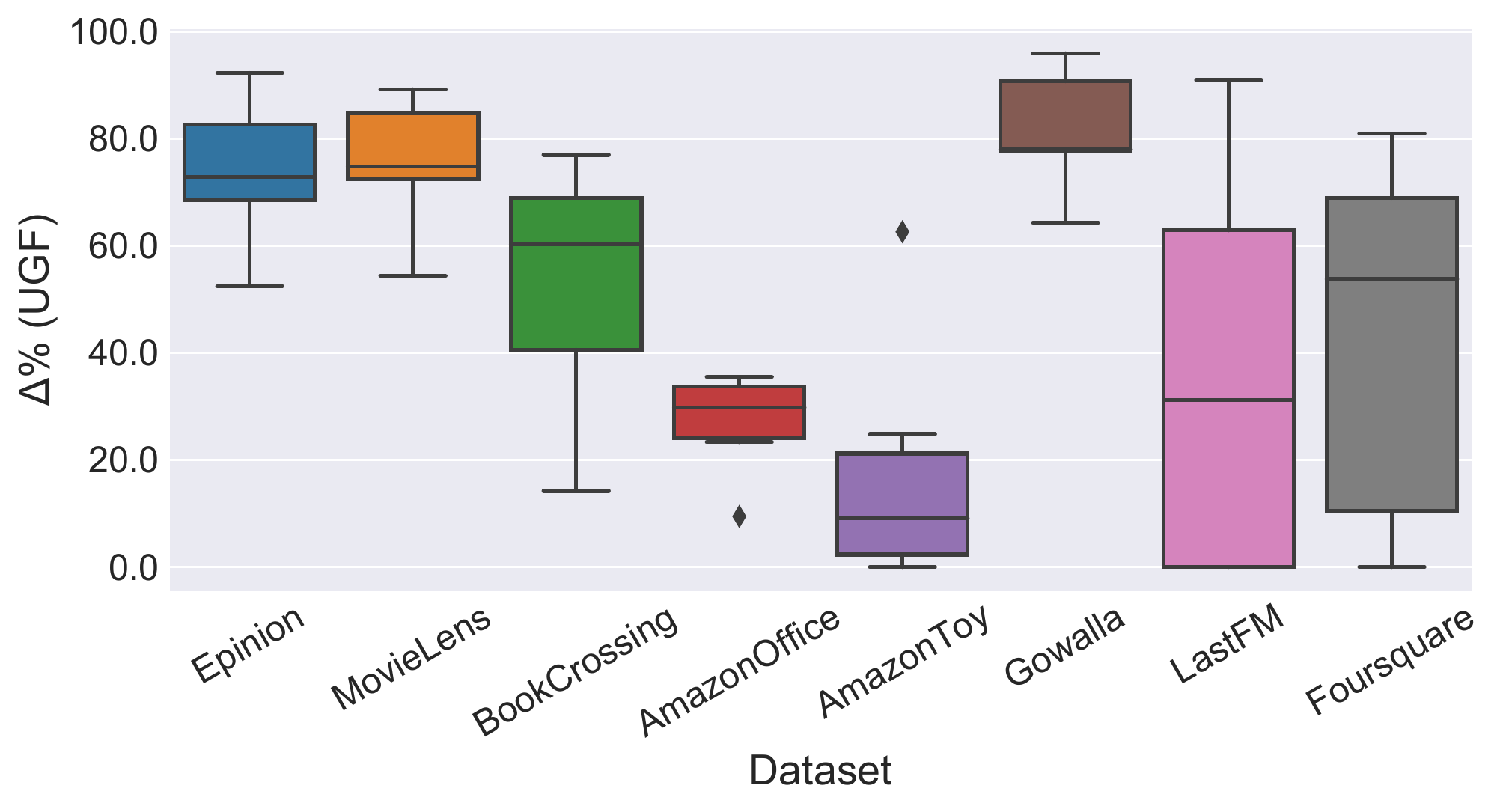}
        \label{fig:impUGF_datasets_top005}%
    }%
    \qquad
    \subfloat[User grouping: G2]
    {
        \includegraphics[scale=0.32]{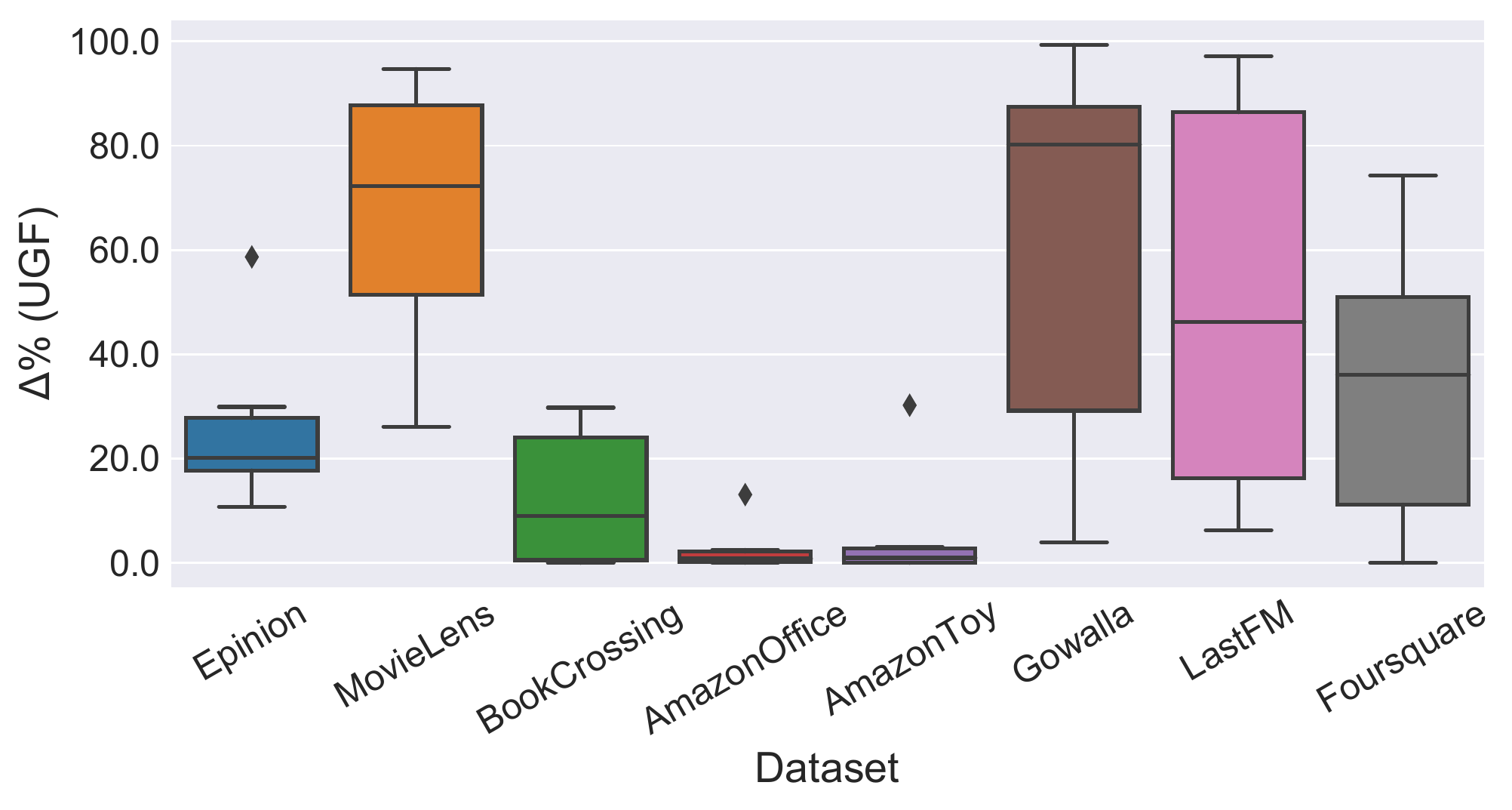} 
        \label{fig:impUGF_datasets_top2}%
    }%
    \caption{Comparison of \model performance in terms of $\Delta\%$UGF improvement across different domains.}%
    \label{fig:impUGF_datasets}%
\end{figure}

\partitle{Effect on UGF.}
In Figure \ref{fig:impUGF_datasets}, we depict \model's performance in terms of UGF improvement ($\Delta\%$(UGF)) on each domain for both user grouping methods. Figure \ref{fig:impUGF_datasets_top005} shows that certain domains exhibit different patterns. It is clear from this figure that mitigating discrimination against the disadvantaged user group can be a challenging task on specific domains. For example, we see that eCommerce domains (Amazon datasets) indicate a lower UGF improvement. We also observe a relatively better performance for POI recommendation datasets (Gowalla and Foursquare). Surprisingly, in Figure \ref{fig:impUGF_datasets_top2} we observe a totally different behavior on some datasets such as Epinions, Gowalla, and BookCrossing depending on the user grouping (G1 vs.~G2). Moreover, eCommerce domains have considerably similar behavior on both G1 and G2. In general, according to Figures \ref{fig:impUGF_datasets_top005} and \ref{fig:impUGF_datasets_top2}, we observe a wider range of variance in UGF improvement on implicit-feedback datasets compared to explicit-feedback datasets. We can argue for this particular observation from two specific points of view, including (1) user grouping methods (interactions vs.~popular consumption) (2) the characteristics of datasets. Next, we analyze the second point of view while we investigate the first one in Section \ref{sec:usergroups}.

\begin{figure}%
    \centering
    \subfloat[Users are categorized by G1.]
    {
        \includegraphics[scale=0.43]{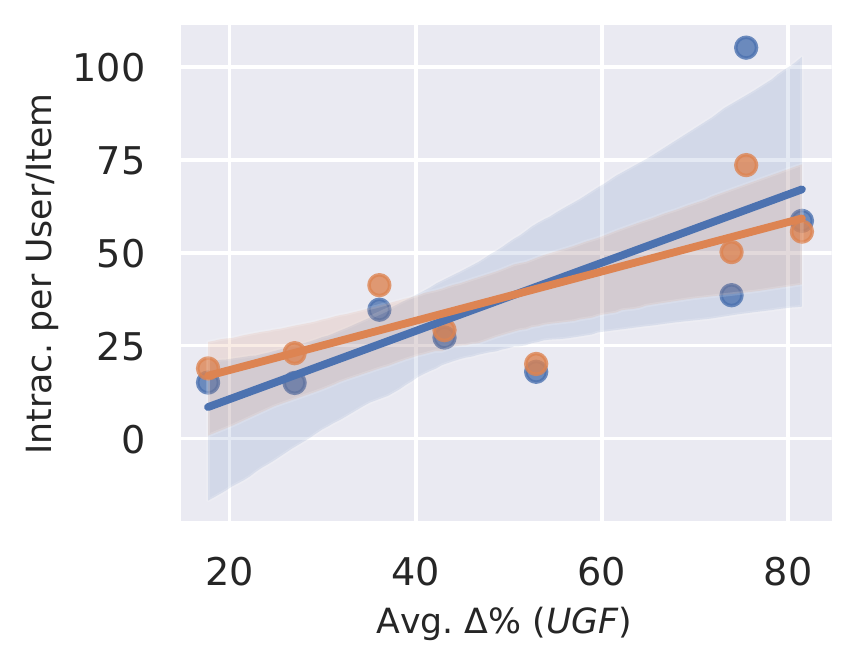}
        \label{fig:dataset_characteristics_ugf_005}%
    }%
    \qquad
    \subfloat[Users are categorized by G2.]
    {
        \includegraphics[scale=0.43]{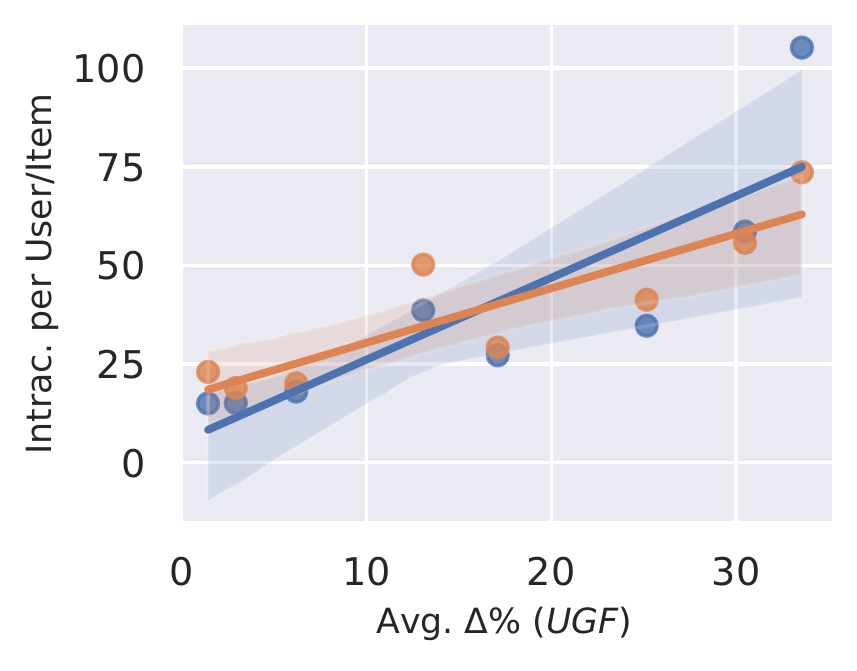} 
        \label{fig:dataset_characteristics_ugf_2}%
    }%
    \caption{Correlations of data characteristics vs.~average $\Delta\%$UGF improvement on all datasets. Blue and orange lines are associated with the number of interactions per user and item, respectively (best viewed in color).}%
    \label{fig:dataset_characteristics_ugf}%
\end{figure}

\partitle{Data characteristics impact on UGF improvement.} 
In Figure \ref{fig:dataset_characteristics_ugf}, for each grouping method we depict a scatter plot in which the horizontal axis (x-axis) is average UGF improvement over all base ranking models on a specific dataset and the vertical axis (y-axis) can be either $\frac{\left| \mathcal{P} \right|}{\left| \mathcal{U} \right|}$ or $\frac{\left| \mathcal{P} \right|}{\left| \mathcal{I} \right|}$  (see Table \ref{tbl:datasets}). In these plots, \textcolor{blue}{blue} lines are concerned with relation between $\frac{\left| \mathcal{P} \right|}{\left| \mathcal{U} \right|}$ and average UGF improvement. The correlation coefficients and p-values of these two variables for interactions and popular consumption grouping methods are ($0.72$, $0.0431$) and ($0.84$, $0.0076$), respectively suggesting a statistically significant correlation. In other words, the higher number of interactions per user can lead to better performance of fairness model. Furthermore, in these plots, \textcolor{orange}{orange} lines represent relation between $\frac{\left| \mathcal{P} \right|}{\left| \mathcal{I} \right|}$ and average UGF improvement. Interestingly, the same behavior exists when investigating the correlation between these two variables. That is, we observe the correlation coefficient and p-values of (0.8076, 0.0153) and (0.8756, 0,0044) for G1 and G2, respectively. Therefore, as the contributions of items in interactions increases, the average improvement of UGF increases. It is worth mentioning that Gowalla and MovieLens have the best average improvement of UGF among implicit and explicit datasets, respectively, since they have the highest values of $\frac{\left| \mathcal{P} \right|}{\left| \mathcal{U} \right|}$ and $\frac{\left| \mathcal{P} \right|}{\left| \mathcal{I} \right|}$ (see Table \ref{tbl:datasets}). Moreover, as Figure \ref{fig:impUGF_datasets} shows, eCommerce domains (Amazon datasets) have the lowest improvement in terms of UGF for both G1 and G2. The reason can be rooted in some data characteristics. For instance, these domains have the lowest values of $\frac{\left| \mathcal{P} \right|}{\left| \mathcal{U} \right|}$ and $\frac{\left| \mathcal{P} \right|}{\left| \mathcal{I} \right|}$. Also, it is worth mentioning that we observe the highest sparsity among all the datasets, suggesting that it can also play a role in our observation.

\begin{figure}%
    \centering
    \subfloat[NDCG]{
        \includegraphics[scale=0.35]{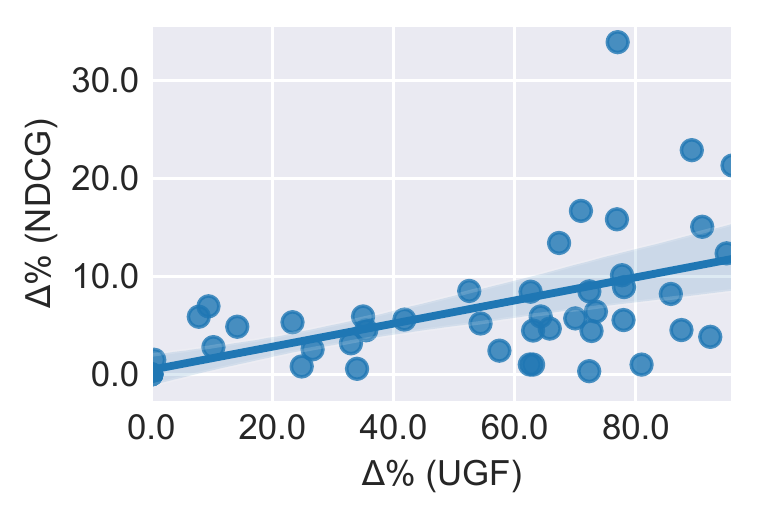}
        \label{fig:regression_ndcg_ugf_005}
    }%
    \subfloat[Novelty]{
        \includegraphics[scale=0.35]{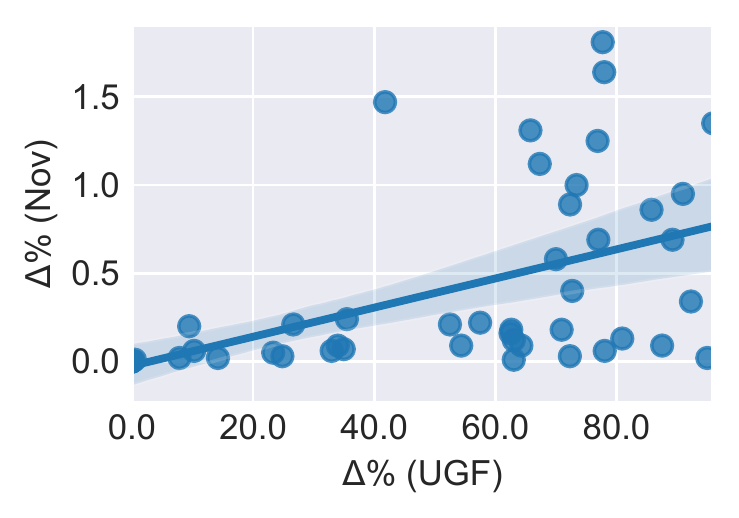}
        \label{fig:regression_nov_ugf_005}
    }%
    \subfloat[$\Delta$GAP]{
        \includegraphics[scale=0.35]{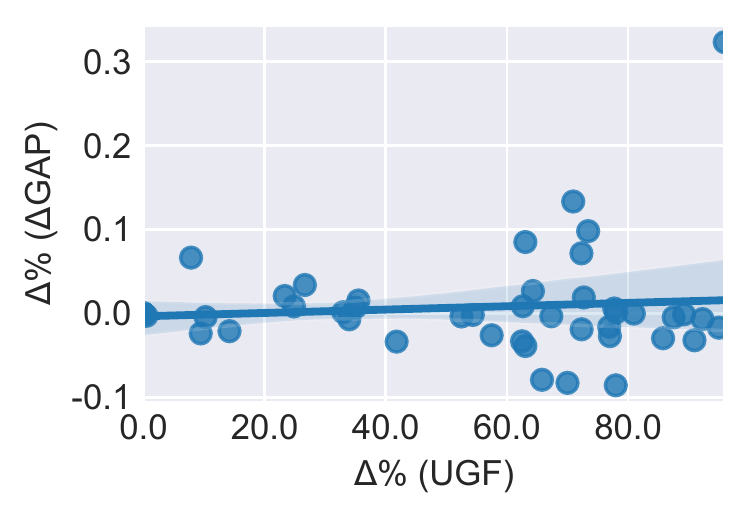}
        \label{fig:regression_gap_ugf_005} 
    }%
    \qquad
    \subfloat[NDCG]{
        \includegraphics[scale=0.35]{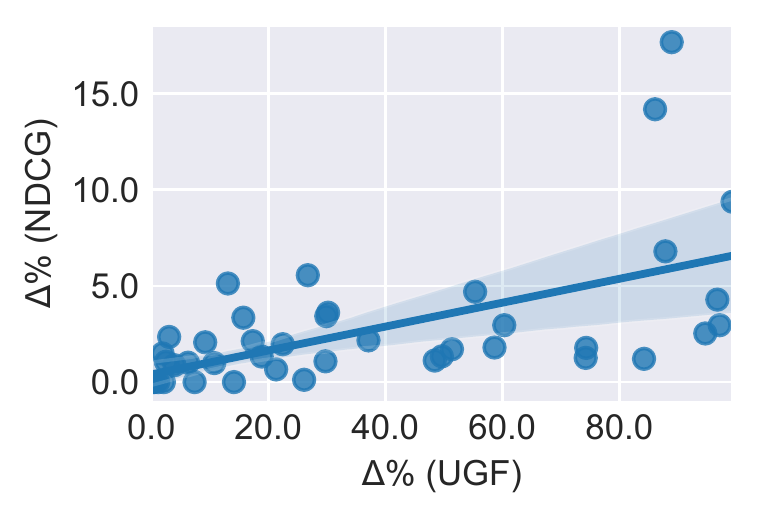}
        \label{fig:regression_ndcg_ugf_20}
    }%
    \subfloat[Novelty]{
        \includegraphics[scale=0.35]{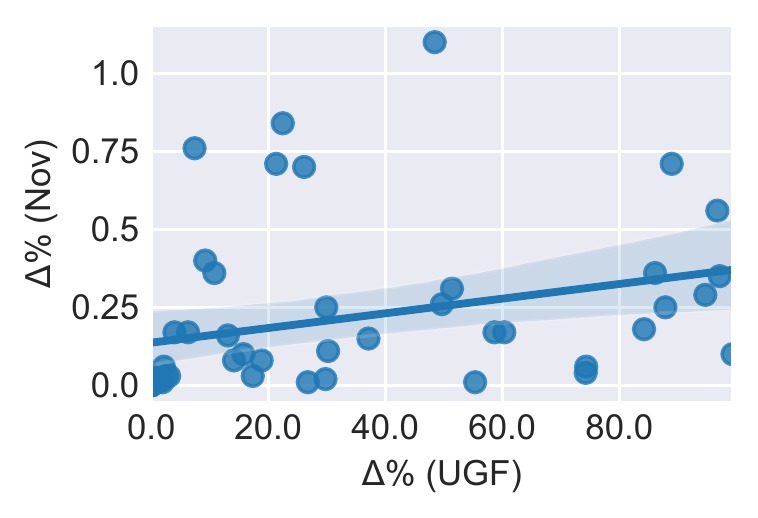}
        \label{fig:regression_nov_ugf_20}
    }%
    \subfloat[$\Delta$GAP]{
        \includegraphics[scale=0.35]{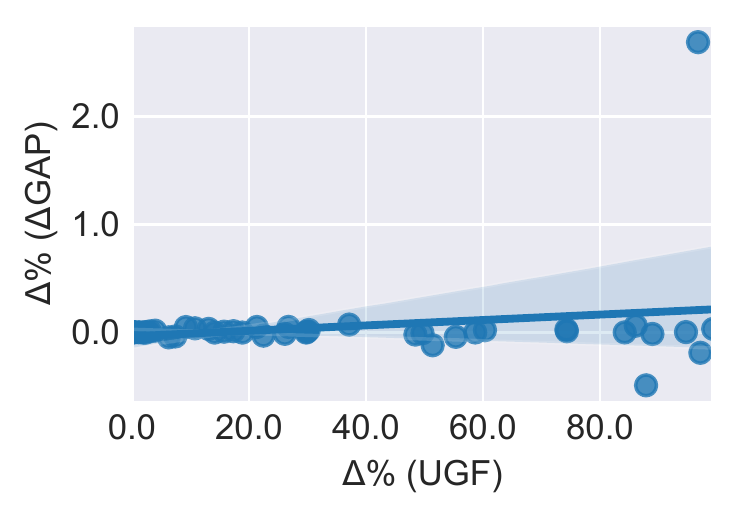}
        \label{fig:regression_gap_ugf_20}
    }%
    \caption{Correlation between \model's improvement in terms of $\Delta\%$UGF and other effectiveness metrics.}%
    \label{fig:regression_plot}%
\end{figure}

\subsection{Base Recommendation Algorithms}
\label{sec:recmodels}
In this experiment, we intend to examine how different types of base ranking models (\eg deep vs.~shallow) treat different user groups before and after applying \model.

\partitle{Base ranking fairness analysis.}
Table \ref{tbl:epinions} and Table \ref{tbl:lastfm} show the detailed results of different base ranking models in which Org. and Fair rows indicate results before and after applying \model, respectively. Here, we intend to investigate which base ranking models are less prone to unfairness before applying \model. According to these tables, almost all base ranking models suffer from the user-fairness issue. MostPop appears to be particularly prone to fairness issues since the most discriminatory treatment toward the disadvantaged user group can be seen in this model. Numerically, this model has the maximum average UGF over all datasets before applying \model. The reason behind this unfair behavior could be the fact that MostPop only recommends the most popular items. Tables \ref{tbl:epinions} and \ref{tbl:lastfm} show that MostPop model does not provide users with any long-tail items in the recommendation results before applying \model. Therefore, the disadvantaged user group is more likely to get worse recommendations since, usually, they interact with more long-tail items in their profile~\cite{rahmani2022unfairness}. On the other hand, PF considers the tastes of both user groups better than other models.
As one can see in Org.~rows of Tables~\ref{tbl:epinions} and \ref{tbl:lastfm}, PF puts more emphasis on long-tail items, compared to other base ranking models, leading to less effectiveness on the disadvantaged user group.

Although the treatment of different base ranking models can be attributed to their characteristics, our observations suggest that the user grouping assumption also affects the sensitivity of a model to fairness. We refer to Table \ref{tbl:lastfm} where in some cases, base raking models provide the disadvantaged group with the same or better recommendations. In this case, the base ranking models do not discriminate against the disadvantaged group on the Last.fm dataset, and seem not to favor the advantaged group. We observe similar behavior of base ranking models on the Foursquare dataset. We bring a comprehensive analysis of the effect of different user grouping methods on the behavior of base ranking models before and after applying \model in Section \ref{sec:usergroups}. 

\begin{figure}%
    \centering
    \subfloat[G1]
    {
        \includegraphics[scale=0.35]{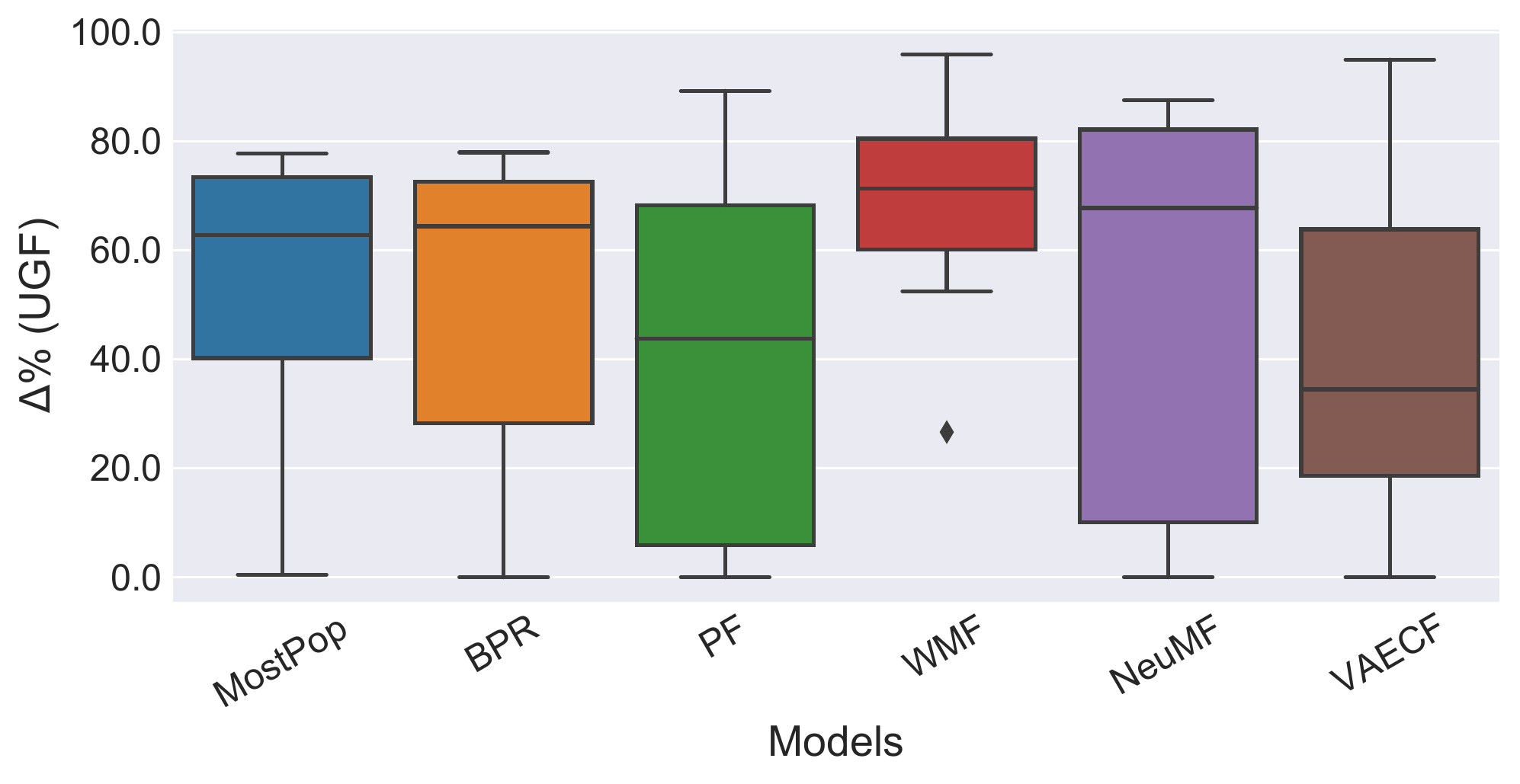}
        \label{fig:ugf_models_top005}
    }%
    \qquad
    \subfloat[G2]
    {
        \includegraphics[scale=0.35]{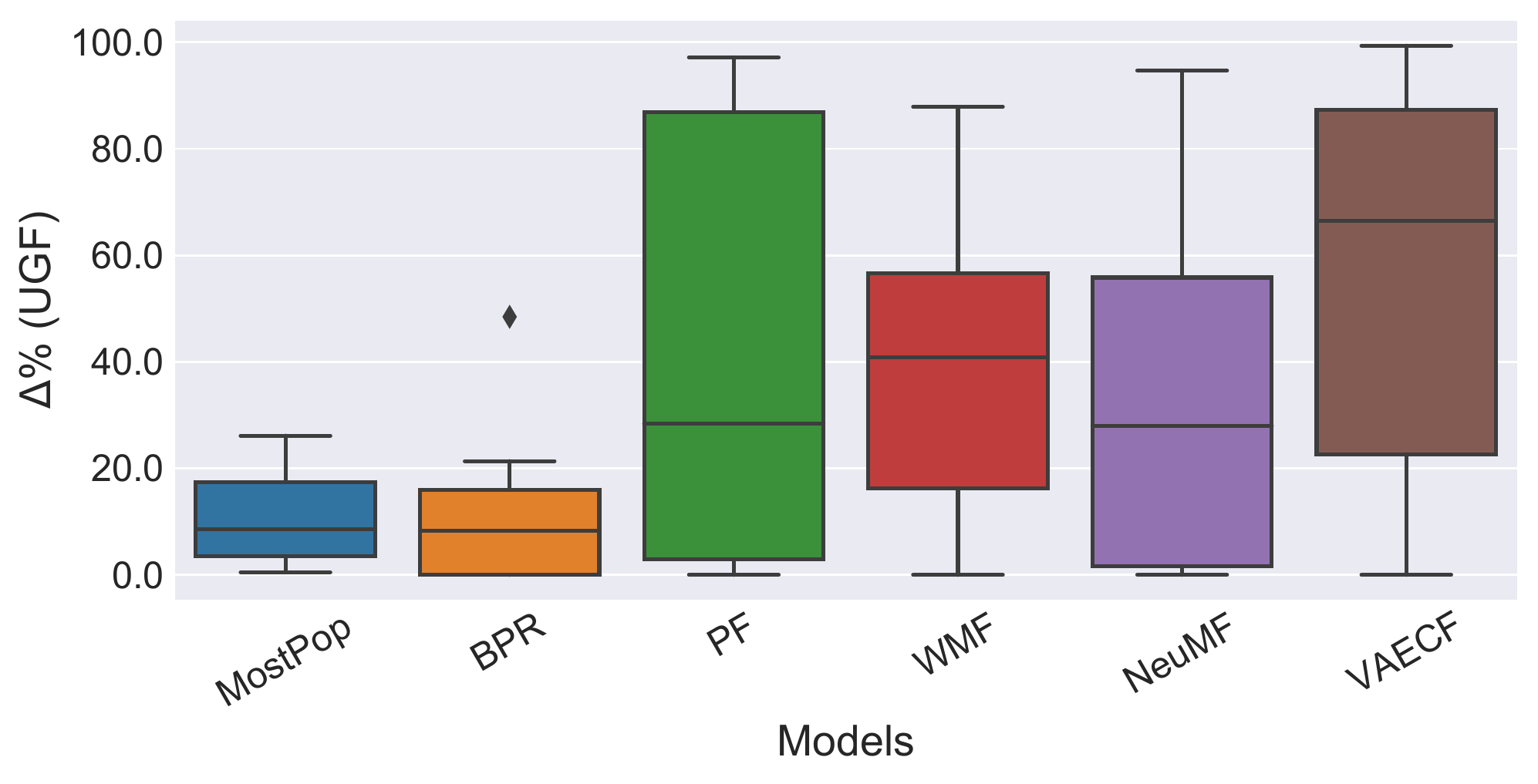} 
        \label{fig:ugf_models_top2}
    }%
    \caption{The \model's performance in terms of $\Delta\%$UGF on different base ranking models.}%
    \label{fig:ugf_models}%
\end{figure}

\partitle{Impact of \model on base ranking models.}
Figure \ref{fig:ugf_models} demonstrates the performance of different models in terms of UGF improvement after applying \model for both grouping methods. We see that post-processing the output of different models using \model exhibits different patterns. Interestingly, Figure \ref{fig:ugf_models_top005} shows that \model manages to mitigate unfairness on the two baseline approaches (\ie MostPop and BPR) much more effectively than VAECF which is a deep recommendation model. Also, they provide comparable results in terms of UGF improvement in comparison to another deep recommendation model (\ie NeuMF). Moreover, MostPop and BPR have a moderate range of variance indicating the fact that they can be considered as robust models when it comes to dealing with different domains. Also, \model applied to WMF demonstrates the most robust performance in mitigating user-oriented unfairness, exhibiting the least variance, as well as the best average improvement in terms of UGF (\ie $68.54\%$). 
On the other hand, Figure \ref{fig:ugf_models_top2} demonstrates the performance of different base ranking models in terms of UGF improvement for G2.

Comparing Figures \ref{fig:ugf_models_top005} and \ref{fig:ugf_models_top2}, we observe two considerably different behaviors of different base ranking models when it comes to the variance and UFG improvement. Specifically, we see that the two baseline approaches (\ie MostPop and BPR) have the worst performance in terms of UGF improvement.  Moreover, VAECF model has two opposite treatments on G1 and G2. This model has the minimum and maximum average improvement of UGF (\ie $41.9\%$ and $55.52\%$) for G1 and G2, respectively. Also, there is a wider range of variance in VAECF model for G2. However, WMF model seems to be the most reliable base ranking model since it has a reasonable variance range and also a considerable average improvement of UGF for both G1 and G2. To sum up, we see that the way advantaged/disadvantaged users are defined (\ie user grouping method) can lead to drastically different results, reiterating the importance of a comprehensive evaluation that considers multiple assumptions.

\begin{table*}
\centering
\caption{The recommendation performance of all, advantaged, and disadvantaged users of \model and corresponding baselines on Epinions datasets for G1 and G2. All re-ranking results here are obtained under the fairness constraint on NDCG. The evaluation metrics here are calculated based on the top-10 predictions in the test set.}
\label{tbl:epinions}
\resizebox{.9\textwidth}{!}{
    \begin{tabular}{lllllllllllllll}
\toprule
& \multirow{2}{*}{Model} & \multirow{2}{*}{Type} & \multicolumn{5}{c}{\textbf{User Relevance (NDCG)}} && \multicolumn{6}{c}{\textbf{Item Exposure}} \\
\cmidrule{4-8} \cmidrule{10-15}
                    &    &                       & All & Adv. & Disadv. & UGF $\downarrow$ & $\Delta$\% && Nov. & Cov. & Short. & Long. & $\Delta$GAP\textsubscript{A}$\downarrow$ & $\Delta$GAP\textsubscript{D}$\downarrow$ \\
\midrule

\multirow{12}{*}{\rotatebox[origin=c]{90}{\textbf{G1}}} & \multirow{2}{*}{MostPop} & Org. & 0.0302 & 0.0566 & 0.0288 & 0.0278 & -- && 3.7133 & 0.49 & 26770.0 & 0.0 & 255.307 & 204.721 \\
                       & & Fair & 0.0301 & 0.0374 & 0.0298 & 0.0077 & 72.3 && 3.7123 & 2.43 & 26770.0 & 0.0 & 287.94 & 205.039 \\ 
                       \cmidrule{2-15}
& \multirow{2}{*}{BPR} & Org. & 0.0297 & 0.055 & 0.0283 & 0.0267 & -- && 3.703 & 0.63 & 26770.0 & 0.0 & 257.095 & 206.271 \\
                       & & Fair & 0.0316 & 0.0384 & 0.0312 & 0.0071 & 73.41 && 3.6658 & 1.55 & 26770.0 & 0.0 & 291.545 & 217.278 \\
                       \cmidrule{2-15}
& \multirow{2}{*}{PF} & Org. & 0.0321 & 0.0751 & 0.0298 & 0.0453 & -- && 4.9602 & 50.73 & 22169.0 & 4601.0 & 83.502 & 61.209 \\
                       & & Fair & 0.0364 & 0.0504 & 0.0356 & 0.0148 & 67.33 && 5.016 & 51.12 & 21872.0 & 4898.0 & 90.315 & 53.907 \\
                       \cmidrule{2-15}
& \multirow{2}{*}{WMF} & Org. & 0.0235 & 0.0577 & 0.0217 & 0.036 & -- && 4.9483 & 29.13 & 25225.0 & 1545.0 & 72.469 & 41.703 \\
                       & & Fair & 0.0255 & 0.0417 & 0.0246 & 0.0171 & 52.5 && 4.9588 & 27.09 & 25335.0 & 1435.0 & 74.201 & 39.559 \\
                       \cmidrule{2-15}
& \multirow{2}{*}{NeuMF} & Org. & 0.0451 & 0.0865 & 0.0429 & 0.0436 & -- && 3.8923 & 12.14 & 26533.0 & 237.0 & 195.495 & 196.681 \\
                       & & Fair & 0.0488 & 0.0548 & 0.0485 & 0.0062 & 85.78 && 3.9259 & 12.62 & 26531.0 & 239.0 & 191.713 & 188.855 \\
                       \cmidrule{2-15}
& \multirow{2}{*}{VAECF} & Org. & 0.0444 & 0.0801 & 0.0425 & 0.0376 & -- && 4.4061 & 31.07 & 24602.0 & 2168.0 & 123.222 & 123.408 \\
                       & & Fair & 0.0461 & 0.0488 & 0.0459 & 0.0029 & 92.29 && 4.4212 & 31.12 & 24550.0 & 2220.0 & 124.084 & 120.874 \\

                       \midrule \midrule
                       
\multirow{12}{*}{\rotatebox[origin=c]{90}{\textbf{G2}}} & \multirow{2}{*}{MostPop} & Org. & 0.0302 & 0.0465 & 0.0261 & 0.0204 & -- && 3.7133 & 0.49 & 26770.0 & 0.0 & 202.723 & 207.972 \\
&                       & Fair & 0.0299 & 0.0445 & 0.0262 & 0.0182 & 10.78 && 3.7 & 0.83 & 26770.0 & 0.0 & 217.749 & 208.241 \\  \cmidrule{2-15}
&\multirow{2}{*}{BPR} & Org. & 0.0304 & 0.0446 & 0.0268 & 0.0178 & -- && 3.708 & 0.68 & 26770.0 & 0.0 & 203.427 & 208.685 \\
&                      & Fair & 0.0302 & 0.0414 & 0.0274 & 0.014 & 21.35 && 3.6818 & 0.87 & 26770.0 & 0.0 & 219.547 & 213.101 \\  \cmidrule{2-15}
&\multirow{2}{*}{PF} & Org. & 0.0321 & 0.0516 & 0.0272 & 0.0244 & -- && 4.9602 & 50.73 & 22169.0 & 4601.0 & 70.422 & 60.071 \\
&                       & Fair & 0.0332 & 0.0469 & 0.0298 & 0.0171 & 29.92 && 4.9727 & 50.92 & 22067.0 & 4703.0 & 73.278 & 57.827 \\  \cmidrule{2-15}
&\multirow{2}{*}{WMF} & Org. & 0.0235 & 0.0392 & 0.0196 & 0.0196 & -- && 4.9483 & 29.13 & 25225.0 & 1545.0 & 45.145 & 42.495 \\
&                      & Fair & 0.024 & 0.0369 & 0.0207 & 0.0162 & 17.35 && 4.9468 & 28.01 & 25290.0 & 1480.0 & 46.865 & 41.703 \\  \cmidrule{2-15}
&\multirow{2}{*}{NeuMF} & Org. & 0.0451 & 0.0663 & 0.0398 & 0.0265 & -- && 3.8923 & 12.14 & 26533.0 & 237.0 & 170.13 & 203.371 \\
&                       & Fair & 0.0457 & 0.0629 & 0.0414 & 0.0215 & 18.87 && 3.8956 & 12.33 & 26542.0 & 228.0 & 170.671 & 202.121 \\  \cmidrule{2-15}
&\multirow{2}{*}{VAECF} & Org. & 0.0444 & 0.0601 & 0.0405 & 0.0196 & -- && 4.4061 & 31.07 & 24602.0 & 2168.0 & 112.638 & 126.135 \\
&                       & Fair & 0.0436 & 0.05 & 0.042 & 0.0081 & 58.67 && 4.4136 & 31.02 & 24576.0 & 2194.0 & 114.078 & 124.504 \\ 

\bottomrule
\end{tabular}
}
\end{table*}

\begin{table*}
\centering
\caption{The recommendation performance of all, advantaged, and disadvantaged users of \model and corresponding baselines on Last.fm datasets for G1 and G2. All re-ranking results here are obtained under the fairness constraint on NDCG. The evaluation metrics here are calculated based on the top-10 predictions in the test set.}
\label{tbl:lastfm}
\resizebox{.9\textwidth}{!}{
    \begin{tabular}{lllllllllllllll}
    \toprule
    & \multirow{2}{*}{Model} & \multirow{2}{*}{Type} & \multicolumn{5}{c}{\textbf{User Relevance (NDCG)}} && \multicolumn{6}{c}{\textbf{Item Exposure}} \\
    \cmidrule{4-8} \cmidrule{10-15}
                        &    &                       & All & Adv. & Disadv. & UGF $\downarrow$ & $\Delta$\% && Nov.  & Cov.  & Short. & Long. & $\Delta$GAP\textsubscript{A}$\downarrow$ & $\Delta$GAP\textsubscript{D}$\downarrow$ \\
    \midrule

    \multirow{12}{*}{\rotatebox[origin=c]{90}{\textbf{G1}}} & \multirow{2}{*}{MostPop} & Org. & 0.0297 & 0.032 & 0.0296 & 0.0024 & -- && 3.3842 & 0.66 & 17970.0 & 0.0 & 53.463 & 111.297 \\
                       && Fair & 0.03 & 0.0309 & 0.03 & 0.0009 & 62.5 && 3.3789 & 0.86 & 17970.0 & 0.0 & 50.013 & 109.27 \\ \cmidrule{2-15}
    &\multirow{2}{*}{BPR} & Org. & 0.0299 & 0.0344 & 0.0297 & 0.0046 & -- && 3.392 & 0.73 & 17970.0 & 0.0 & 31.908 & 81.347 \\
                           && Fair & 0.0302 & 0.0319 & 0.0302 & 0.0017 & 63.04 && 3.3917 & 1.19 & 17970.0 & 0.0 & 41.885 & 81.01 \\ \cmidrule{2-15}
    &\multirow{2}{*}{PF} & Org. & 0.0372 & 0.0259 & 0.0378 & -0.0119 & -- && 5.0905 & 66.82 & 12480.0 & 5490.0 & -50.942 & -23.749 \\
                           && Fair & 0.0372 & 0.0259 & 0.0378 & -0.0119 & 0.0 && 5.0905 & 66.82 & 12480.0 & 5490.0 & -50.942 & -23.749 \\ \cmidrule{2-15}
    &\multirow{2}{*}{WMF} & Org. & 0.0319 & 0.0498 & 0.0309 & 0.0188 & -- && 5.536 & 78.04 & 8000.0 & 9970.0 & -50.864 & -45.98 \\
                           && Fair & 0.0367 & 0.0383 & 0.0366 & 0.0017 & 90.96 && 5.5886 & 74.52 & 7721.25 & 10248.75 & -42.613 & -51.132 \\ \cmidrule{2-15}
    &\multirow{2}{*}{NeuMF} & Org. & 0.0415 & 0.0394 & 0.0416 & -0.0022 & -- && 3.7975 & 9.16 & 17863.0 & 107.0 & 31.44 & 69.969 \\
                           && Fair & 0.0415 & 0.0394 & 0.0416 & -0.0022 & 0.0 && 3.7975 & 9.16 & 17863.0 & 107.0 & 31.44 & 69.969 \\ \cmidrule{2-15}
    &\multirow{2}{*}{VAECF} & Org. & 0.056 & 0.0651 & 0.0556 & 0.0095 & -- && 4.604 & 42.93 & 13890.0 & 4080.0 & -6.103 & 10.94 \\
                           && Fair & 0.056 & 0.0651 & 0.0556 & 0.0095 & 0.0 && 4.604 & 42.93 & 13890.0 & 4080.0 & -6.103 & 10.94 \\
                           
    \midrule \midrule
    
    \multirow{12}{*}{\rotatebox[origin=c]{90}{\textbf{G2}}} & \multirow{2}{*}{MostPop} & Org. & 0.0297 & 0.0564 & 0.023 & 0.0334 & -- && 3.3842 & 0.66 & 17970.0 & 0.0 & 26.813 & 146.477 \\
    &                       & Fair & 0.0294 & 0.0544 & 0.0231 & 0.0313 & 6.29 && 3.3784 & 0.86 & 17970.0 & 0.0 & 20.784 & 144.465 \\ \cmidrule{2-15}
    &\multirow{2}{*}{BPR} & Org. & 0.029 & 0.0455 & 0.0249 & 0.0206 & -- && 3.3901 & 0.66 & 17970.0 & 0.0 & 3.963 & 102.065 \\
    &                       & Fair & 0.0284 & 0.0433 & 0.0246 & 0.0187 & 9.22 && 3.3767 & 1.13 & 17970.0 & 0.0 & 9.943 & 101.255 \\ \cmidrule{2-15}
    &\multirow{2}{*}{PF} & Org. & 0.0372 & 0.0485 & 0.0344 & 0.0141 & -- && 5.0905 & 66.82 & 12480.0 & 5490.0 & -39.191 & -19.012 \\
    &                       & Fair & 0.0361 & 0.0364 & 0.036 & 0.0004 & 97.16 && 5.0726 & 66.62 & 12538.0 & 5432.0 & -24.996 & -22.189 \\ \cmidrule{2-15}
    &\multirow{2}{*}{WMF} & Org. & 0.0319 & 0.0453 & 0.0285 & 0.0168 & -- && 5.536 & 78.04 & 8000.0 & 9970.0 & -52.414 & -43.367 \\
    &                       & Fair & 0.0334 & 0.0394 & 0.0319 & 0.0075 & 55.36 && 5.5355 & 75.25 & 7996.0 & 9974.0 & -45.46 & -46.353 \\ \cmidrule{2-15}
    &\multirow{2}{*}{NeuMF} & Org. & 0.0415 & 0.0682 & 0.0348 & 0.0334 & -- && 3.7975 & 9.16 & 17863.0 & 107.0 & 4.283 & 97.967 \\
    &                       & Fair & 0.0406 & 0.0575 & 0.0364 & 0.021 & 37.13 && 3.7919 & 9.42 & 17857.0 & 113.0 & 13.308 & 96.281 \\ \cmidrule{2-15}
    &\multirow{2}{*}{VAECF} & Org. & 0.056 & 0.0808 & 0.0498 & 0.0309 & -- && 4.604 & 42.93 & 13890.0 & 4080.0 & -14.901 & 21.744 \\
    &                       & Fair & 0.0536 & 0.0544 & 0.0534 & 0.001 & 96.76 && 4.578 & 43.13 & 13935.3333 & 4034.6667 & 4.642 & 20.596 \\
    \bottomrule
    \end{tabular}
}
\end{table*}
\vspace{-5pt}
\subsection{User Groups Assumptions}
\label{sec:usergroups}
In this section, we examine the effect of the user grouping methods on the performance of the \model. In Section \ref{sec:fairness_assumption}, we introduce two common settings that are used by previous studies \cite{abdollahpouri2019unfairness,li2021user} to group users according to their level of activity and the consumption of popular items. Therefore, herein, the main question is to what extent the user grouping assumption could impact the performance or fairness between different user groups.

\partitle{Effect of user grouping method on performance of \model.} 
In this work, we aim to study the performance of \model, under other user group assumptions that are not studied in the original paper, enabling us to compare the improvements with other works in the literature and examine the effectiveness of the model when different assumptions are taken into account.
To this end, we consider the following user grouping techniques, namely, G1 and G2 (see Section \ref{sec:fairness_assumption}). Figure \ref{fig:impUGF_datasets} depicts \model's performance across different datasets evaluated by $\Delta\%$UGF for G1 and G2. \model enhances recommendation fairness on both G1 and G2 across different datasets. However, we see apparent different performances on G1 and G2. For instance, \model achieves higher performance in Epinions on G1 compared to G2. Thus, the main question is the reason for different treatment of \model when grouping users differently. Comparing Figures \ref{fig:impUGF_datasets_top005} and \ref{fig:impUGF_datasets_top2}, we observe a worse performance on G2 compared to G1. This indicates that the higher the number of users of the advantaged group, the more challenging the task would be (see Table \ref{tbl:datasets}). Moreover, we see that the box plots show a wider range of variance in terms of the performance of different models, suggesting that the tested models are less robust when it comes to dealing with more users. Interestingly, we see a very large drop in the performance in Epinions.
Looking at Table \ref{tbl:epinions}, we observe that \model's UGF improvement shows a high correlation with the performance difference between the two groups in the original ranking, that is, the more unfair an original ranking is, the more UGF improvement is achieved after applying \model (Pearson's $r = 0.36185 $, $p < 0.05$). This shows a clear dependence of the achieved improvement on how users are grouped.

Moreover, we see a reverse behavior in Last.fm, where the performance with G2 is better than the performance with G1. This finding is in line with the study of \citet{kowald2020unfairness} and the reason could be related to the fact that the users on Last.fm have relatively more average interactions, compared to other datasets. Therefore, considering only top-$5\%$ of users with the largest profile size as an advantaged group can not represent the unfairness treatment of fairness-unaware recommendation algorithms. As Table \ref{tbl:lastfm} shows, the recommendations provided by most base ranking models fairly represent the tastes of both groups of users. On the other hand, as can be seen in Table \ref{tbl:lastfm}, G2 can reveal the unfairness issue on the  user side on Last.fm.
This observation can determine that G2 is not influenced by user profile size as it differentiates users according to the number of popular items in their profiles, irrespective of their profile size. Consequently, based on our experiments, G2 appears to be a relatively more generalizable grouping method than G1.

To sum up, the instability of results on different datasets can be rooted in the fact that classifying users based on a cutoff (\eg top 5\% users based on the number of interactions) is not an accurate way of classifying advantaged vs.~disadvantaged user groups. Thus, we suggest to group users using approaches that classify users based on multiple features such as gender, age, race, and interactions count simultaneously. In general, finding a more stable and reasonable way of classifying users such as K-means (or other clustering techniques) is a potential topic for future work.

\partitle{Impact of user grouping on $\Delta\%$UGF across base ranking models.}
When comparing Figures \ref{fig:ugf_models_top005} and \ref{fig:ugf_models_top2}, we see that various base ranking models show inconsistent behavior in terms of UGF improvement for different grouping methods. Interestingly, we see a large drop in the performance of the MostPop and BPR models with the popular consumption grouping method, G2, in comparison to G1.
Previous studies (\eg \citet{mansoury2020feedback}) show that BPR favors popular items and gives them a high rank in  the recommendation list. Clearly, the same holds for MostPop algorithm. We also see a drop, to a lower extent, in the performance of WMF, and NeuMF in $\Delta\%$UGF for G2. On the other hand, PF and VAECF models show a more robust behavior with changes in the grouping assumption while the latter (VAECF), achieves slightly better performance (see Figure \ref{fig:ugf_models}). Our experiments show that this variability is explainable by the extent different algorithms are prone to popularity bias, \ie base ranking models that mostly include popular items in the recommended list are less stable in the improvement of user-oriented fairness ($\Delta\%$UGF) with changes in the grouping method. For example, the average number of long-tail items in the original top-10 recommendation list in BPR, MostPop, PF, WMF, NeuMF, and VAECF across all datasets are 0, 0, 5083, 6211.87, 157, and 3102.75, respectively. This shows that with the exception of WMF, all algorithms with a small number of long-tail items in the original recommendation list suffer from this instability in the improvement of fairness.

\subsection{Fairness vs.~Effectiveness Metrics}
The main objective of this aspect is to mine the underlying trends and trade-offs between various evaluation metrics in recommender systems~\cite{rahmani2022unfairness}. In particular, we are interested in analyzing the potential correlations and trade-offs between UGF improvement, \ie mitigating consumer-oriented fairness, and improvement of other effectiveness metrics such as overall accuracy (NDCG), providers' (items') exposure, the novelty of recommendation, and popularity bias among user groups. Figure \ref{fig:regression_plot} shows the regression plot of overall datasets and models with G1 and G2. Each point represents one run of \model with a baseline model in one domain. The horizontal axis (x-axis) shows the UGF improvement (\ie $\Delta \%$UGF) of that run while the vertical axis (y-axis) denotes the improvement in terms of another effectiveness metric, namely, NDCG, Novelty, and $\Delta\%$GAP. In what follows we discuss these correlations separately.

\partitle{Correlation between $\Delta\%$UGF and $\Delta\%$NDCG.}
Surprisingly, as can be seen in Figures \ref{fig:regression_ndcg_ugf_005} and \ref{fig:regression_ndcg_ugf_20}, there exists a positive correlation between improvement in terms of UGF (\ie $\Delta \%$UGF) and improvement in the overall accuracy of the recommender system measured by NDCG. The Pearson's $r$ correlation coefficient and $p$-values are (0.55036, 5.06e-05), (0.60236, 5.9e-06) respectively for G1 and G2, suggesting a moderate uphill (positive) relationship. In other words, improving user-oriented fairness by applying \model does not necessarily sacrifice the overall accuracy of the system. The reason lies behind the fact that the total NDCG of the recommender system is closely dependent on NDCG of disadvantaged users as they are the majority and \model tries to achieve fairness by giving more attention to disadvantaged users leading to an increase in total accuracy. Across all our experiments, we have seen an average 41.73\% improvement over UGF and 4.446\% over NDCG. 

\partitle{Correlation between $\Delta\%$UGF and $\Delta\%$Nov.}
Figures \ref{fig:regression_nov_ugf_005} and \ref{fig:regression_nov_ugf_20} show the correlation between $\Delta \%$UGF and the novelty of the recommender system (Nov). The Pearson's $r$ correlation coefficients and $p$-values are (0.512, 0.0001994) and (0.30056, 0.037921) for G1 and G2, respectively. As can be seen, there is a weak uphill correlation (+0.3 correlation) between improvement in fairness and overall novelty of recommender systems and among all experiments, we have seen an average improvement of novelty by 0.304. That is due to the fact that \model mainly improves user-oriented fairness of recommendation by including less popular items on top of the ranking list. Therefore, as the results in Tables \ref{tbl:epinions} and \ref{tbl:lastfm} suggest, we naturally observe more recommended long-tail items, causing an improvement in novelty in the fair ranking. 

\partitle{Correlations between $\Delta\%$UGF and $\Delta\%\Delta$GAP.}
The correlation between these two metrics is illustrated in Figures \ref{fig:regression_gap_ugf_005} and \ref{fig:regression_gap_ugf_20}. The Pearson's $r$ coefficients and $p$-values for G1 and G2 are (0.10307, 0.4857423) and (0.2131, 0.145885) respectively. This suggests that these two evaluation metrics are not meaningfully correlated. The variation in the $\Delta\%$UGF cannot explain the observed variability in $\Delta\%\Delta$GAP, and \model does not affect the existing popularity bias in the dataset. Moreover, our experiment indicates the average values of 97.06 and 91.01 for $\Delta$GAP for G1 and G2, respectively, indicating that the average popularity of items in the recommended list is significantly higher than expected according to user profiles after running \model. Therefore, it is vital to investigate how this model can be extended to incorporate mitigation of popularity bias as well as improvement in UGF.
\section{Conclusions, Implications, and Future Work}
\label{sec:conclusions}
In this paper, we conducted an extensive reproducibility study of the work by \citet{li2021user}. The original work focuses on the fairness in recommendation algorithms from the user perspective, evaluating three eCommerce datasets. In this work, we re-implemented the original work and studied its performance under various conditions, namely, fairness assumption, base ranking model, datasets, and evaluation aspects. We provided a deep analysis of how user-oriented fairness can be affected by different user grouping assumptions and data distributions. Moreover, in this work, we aimed at studying the effect of user-oriented fairness on item exposure. \citet{li2021user} treat recommendation in the eCommerce domain like a one-sided market; however, item exposure is obviously as important as user fairness, as it translates to vendors' revenue. As such, we studied the implications of conducting user-oriented fairness on item exposure and showed how mitigating item fairness can affect item exposure. Our experiments on various reproducibility aspects led to interesting new observations and possible future directions in this area, summarized below.

\partitle{Domain dependency.}
We evaluated the performance of user-oriented fairness on six domains. Our main findings indicate that the effectiveness of \model highly depends on the domain. For instance, we found a significantly higher improvement in user fairness on POI domain than in eCommerce. Furthermore, our experiment shows that this variability in performance can be explained using some data characteristics. In particular, we found that there exists a strong positive correlation between improvement in fairness with both average interaction per user ($\frac{\left| \mathcal{P} \right|}{\left| \mathcal{U} \right|}$) and average interaction per item ($\frac{\left| \mathcal{P} \right|}{\left| \mathcal{I} \right|}$). That is, domains with high value of $\frac{\left| \mathcal{P} \right|}{\left| \mathcal{U} \right|}$ or $\frac{\left| \mathcal{P} \right|}{\left| \mathcal{I} \right|}$ can be improved in terms of fairness to a greater extent.
We conclude that it is crucial to evaluate the fairness models on various domains, since data in different domains can have different types of bias and distributions, such as distinct feedback types.

\partitle{Baseline model dependency.}
Our experiment on reproducing \model using six new base recommendation algorithms shows that \model is algorithm agnostic. In other words, it will improve fairness regardless of the baseline recommendation algorithm. However, the extent of this improvement significantly varies across different algorithms. We conclude that, especially for fairness algorithms that do not assume having access to the relevance judgments, it is useful to evaluate the model on different base ranking models.

\partitle{User grouping assumptions.}
User fairness arises mainly when we can demonstrate that a particular recommendation algorithm does not truly represent the tastes of one user group while recommendations are provided for other groups that are consistent with their preferences. It is, therefore, crucial to determine if a grouping criterion can capture this unfair treatment properly. For instance, we observed that the interactions cutoff (top-5\% users based on the interactions) method proposed \cite{li2021user} does not work on Last.fm dataset, and no unfair behavior is observed on this dataset when grouping is done by this method. Based on the findings of this study, we found that the user grouping method is one of the most important aspects of the user fairness algorithm since it directly affects our interpretation of an algorithm's fair behavior. Furthermore, our result shows that grouping users according to a cutoff (\eg top 5\% users based on the number of interactions) is a domain-dependent method with poor generalizability. Hence, a new research line would be to define a domain-independent method of grouping users on the basis of multiple features simultaneously. 
In addition, exploring other user grouping strategies that can show less dependency on the data distribution and domain properties should be explored in this area. Based on our experiments, we envisage using clustering algorithms (such as K-means) can be a promising future research direction in group fairness problems in recommendation systems.

\partitle{Fairness evaluation by multi-sided metrics.}
In general, through extensive experiments, we conclude that a well-tuned \model model can improve user-oriented fairness among groups considerably by achieving higher overall accuracy. This achievement is without sacrificing producers' exposure or general novelty. However, \model does not have a statistically significant effect on mitigating popularity bias among users with different degrees of interest toward popular items, \ie calibration fairness in recommendation system. It is worthwhile to explore methodologies to add calibration fairness and provider aspect to \model as an extra objective or to add them as additional fairness constraints, thus accounting for the shortcoming and taking into account the needs for a multi-sided marketplace.

Overall, our extensive reproducibility experiments on user-oriented post-processing fairness algorithms revealed several novel observations, suggesting various possible directions in this area, especially on the evaluation of such techniques. In the future, we aim to extend this study to multiple fairness algorithms, each of which focuses on a different aspect of fairness or bias. Our goal is to evaluate the extent of generalizability of our findings in this study and propose a general set of guidelines for a \textit{fair} evaluation of fairness algorithms. We believe that our experiments and shared resources open opportunities for both reproducibility and evaluation in this area.

\bibliographystyle{ACM-Reference-Format}
\bibliography{references}

\end{document}